\documentclass[]{pasj01}

\begin{document}
\Received{}
\Accepted{}

\title{A self-consistent leptonic-hadronic interpretation of the electromagnetic and neutrino emissions from blazar TXS 0506+056 }

\author{Gang \textsc{Cao}\altaffilmark{1}%
}
\altaffiltext{1}{Department of  Mathematics, Yunnan University of Finance and Economics, Kunming 650221, Yunnan, P. R. China}
\email{gcao@ynufe.edu.cn}

\author{Chuyuan \textsc{Yang}\altaffilmark{2}}
\altaffiltext{2}{Yunnan Observatories, Chinese Academy of Sciences, Kunming 650216, Yunnan, P. R. China}

\author{Jianping \textsc{Yang}\altaffilmark{3}}
\altaffiltext{3}{Yunnan Agricultural University, Kunming 650201, Yunnan, P. R. China}

\author{Jiancheng \textsc{wang}\altaffilmark{2}}
\KeyWords{BL Lacertae objects: individual( TXS 0506+056 ) -- galaxies: active -- galaxies: gamma-ray -- neutrino -- radiation mechanisms: non-thermal} 

\maketitle

\begin{abstract}
The potential association between the blazar TXS 0506+056 and the neutrino event IceCube-170922A provides a unique opportunity to study the possible physical connection between the high-energy photons and neutrinos. We explore the correlated electromagnetic and neutrino emissions of blazar TXS 0506+056 by a self-consistent leptonic-hadronic model, taking into account  particle stochastic acceleration and all relevant radiative processes self-consistently. The electromagnetic and neutrino spectra of blazar TXS 0506+056  are  reproduced  by the proton synchrotron and hybrid leptonic-hadronic models based on the proton-photon interactions. It is found that the hybrid leptonic-hadronic model can be used to better explain the observed X-ray and $\gamma$-ray spectra of blazar TXS 0506+056 than the proton synchrotron model. Moreover, the predicted neutrino spectrum of the hybrid leptonic-hadronic model is closer to the observed one compared to the proton synchrotron model. We suggest that the hybrid leptonic-hadronic model is more favored if the neutrino event IceCube-170922A is associated with the blazar TXS 0506+056.
\end{abstract}

\section{Introduction}

Neutrino observation by IceCube has opened up a new window in the study of nonthermal processes in astrophysical objects. However, the sources responsible for the neutrino emission have not been identified so far. As the neutrinos are not absorbed when interacting with the background photons or the matters, they can be detected even though the source is far away.
The observed distribution of their arrival direction suggests a predominantly  extragalactic origin. Extragalactic sources, such as active galactic nuclei \citep{mur14,pet15,pad16,gao17,mur18,gao19}, $\gamma$-ray burst \citep{mur06,pet14} and supernova \citep{mur11,pet17}, have been proposed as the potential high-energy neutrino sources. Blazars are believed to be the most promising candidate sources with high-energy neutrino emission.

The IceCube observation recently reported the detection of a neutrino event( IceCube-170922A ), which is coincidence with the blazar TXS 0506+056 during its flaring state \citep{ice18a,ice18b}.
Following the neutrino alert, the blazar TXS 0506+056 is detected in a multi-wavelength campaign, ranging from the radio to $\gamma$-ray bands \citep{ice18b}. The multi-wavelength observations characterize the polarization, variability and energy spectrum of the blazar TXS 0506+056. This source is also first detected in very high-energy $\gamma$-ray bands with the MAGIC Cherenkov telescopes. A chance coincident of the high-energy neutrino with multi-wavelength flare is rejected at a 3.5 $\sigma$ level. The potential association between the activity of TXS 0506+056 and the neutrino event suggests that this object could be the counterpart of the neutrino event.

Blazars are a subclass of radio-loud active galactic nuclei powered by supermassive black holes.  Their radiation is thought to originate in a relativistic jet oriented at a small angle with respect to the line of sight. Blazars are often classified into  BL Lac objects and flat spectrum radio quasars (FSRQs).
BL Lacs have weak or absent emission lines, while FSRQs usually show strong broad emission lines.
The spectrum energy distributions (SEDs) of blazars are characterized by non-thermal continuum spectra with a broad low-energy component from radio-UV to X-ray and a broad high-energy component from  X-ray to $\gamma$-ray. It is generally accepted that the low-energy component of blazar SEDs is produced by synchrotron emission from relativistic electrons accelerated in the jet of blazar. The high-energy component is often interpreted as the inverse Compton (IC) upscattering of ambient soft photons by the accelerated electrons (e.g., B\"{o}ttcher 2007).
The soft photons can be either synchrotron photons within the jet( the synchrotron self-Compton, SSC, process, Maraschi et al. 1992; Bloom and Marscher 1996), or the photons external to the jet ( the  External Compton, EC, process). These external photons may be the UV accretion disk photons\citep{der93}, the accretion disk photons reprocessed by broad-line region  clouds \citep{sik94}, or infrared photons from the dust torus \citep{bla00}.
The above scenario is called the leptonic model. Such models have achieved great successes in explaining the multi-wavelength emission and variability from blazars \citep{bot02,wei10}.

It is physically plausible that the protons are co-accelerated with the electrons to very-high energy by the same mechanism in the jet of blazar. In the so-called hadronic model, the synchrotron emission from the high-energy proton can dominate the high-energy component in the SED of blazars \citep{man93,muc00}. Moreover, the high-energy protons can interact with the background photons to produce the secondary electrons, the synchrotron emission from the secondary electrons can have a significant contribution to the  high-energy component \citep{pet12,mas13,cer15,wei15,dil16,zec17}. The high-energy component can also be produced  by the $\pi^{0}$ decay from the  proton-photon($p\gamma$) interactions \citep{sah13,cao14}. For a recent review on blazar hadronic modelling, see B\"{o}ttcher et al. (2013).

IceCube observation implied that the blazar TXS 0506+056 could be a high-energy neutrino source. The neutrinos are generally associated with the hadronic processes in the jet of blazar.
High-energy neutrinos can be produced by the decay of charged pions from the $p\gamma$ interaction. Therefore, the neutrinos can be considered as the unique signature of these hadronic interactions.
The recent study revealed that the multi-TeV gamma-rays of blazars can be well explained by the photohadronic process, which provides a strong evidence that the neutrino emission from blazars may originate in the photohadronic process \citep{sah19}.
In this paper, we study the correlated electromagnetic and neutrino emissions of blazar TXS 0506+056 by a self-consistent leptonic-hadronic model, taking into account both  electrons and protons
stochastic acceleration and all relevant radiative processes self-consistently. We reproduce the electromagnetic and neutrino emissions of blazar TXS 0506+056  by the proton synchrotron and hybrid leptonic-hadronic models based on the $p\gamma$ interaction. We demonstrate that the observed neutrino signature can allow us to distinguish these different emission models.

In Section 2 we give a brief description of the model. In Section 3 we apply the model to explain the electromagnetic and neutrino emissions of blazar TXS 0506+056. The discussion and conclusion are presented in Section 4.  Throughout this paper, we adopt the cosmological parameters of $H_0 = 70 $  km  s$^{-1}$  Mpc$^{-1}$, $\Omega_M = 0.3$, $\Omega_{\Lambda} = 0.7$.\\

\section{model}

\subsection{Model geometry }

In this section, we give a brief description of the model introduced by \citet{wei15}.  We improve the model of  \citet{wei15} by implementing the kinetic equation of the neutrinos and the Bethe-Heitler process. For a detail description about the model, see \citet{wei10,wei15}.

The model assumes a spherical geometry with two zones, where a acceleration zone with radius $R_{\rm acc}$ is nested within a larger radiation zone with a radius $R_{\rm rad}$.
Both zones are assumed to be homogeneous and to contain isotropic electron and proton distributions as well as a randomly oriented magnetic field. The considered blob travels down the jet axis towards the observer with a bulk Lorentz factor $\Gamma$,
the upstream material is picked up into the acceleration zone where a highly turbulent zone is formed at the edge of the blob.
Here,  both injected particle species are subjected to stochastic acceleration processes up to the relativistic energies  balanced by their radiation losses. However, the acceleration is assumed to be inefficient in the considerable larger radiation zone. The kinetic equations  for each particle species $i$ in each zone can be derived from the relativistic Vlasov equation (Schlickeiser 2002) by  one-dimensional diffusion approximation using the relativistic approximation $p_{i}=m_{i}c$. Since blazar jets are almost aligned with the line of sight of the
observer, we assume the Doppler factor $\delta\simeq \Gamma$. All calculations are conveniently made in the rest-frame of the blob.\\

\begin{figure*}
\centering
\includegraphics[width=13. cm,height=8. cm]{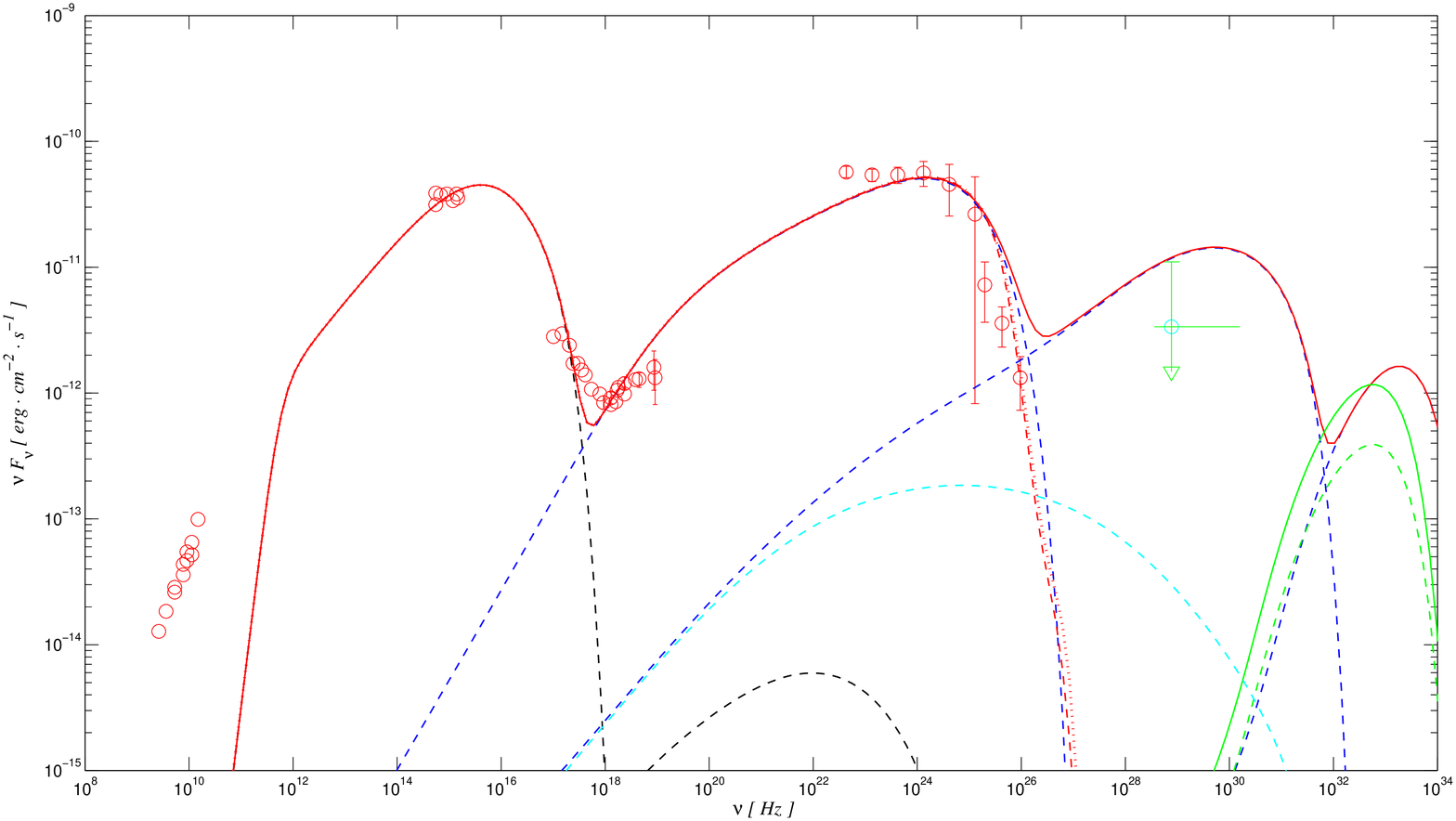}
\caption{Predicted multiwavelength flux and neutrino flux of TXS 0506+056 for the proton synchrotron model. The red circles are the observed multi-wavelength data and the green circle is the detected neutrino flux from IceCube observation. The black dashed curves represent the synchrotron emission and the SSC emission, respectively (from left to right). The blue dashed curves represent the proton synchrotron emission, the synchrotron emission from the secondary pairs  and the $\gamma$-ray emission from $\pi^{0}$ decay, respectively (from left to right).
The cyan dashed curves represent the synchrotron emission of the secondary pairs from the Bethe-Heitler process. The green dashed curve represents the muon neutrino spectrum from the charged pions decay. All flavor neutrino spectrum is also shown as the green solid curve. The red solid curve is the total spectrum from all emission components, while the red dashed curve is the EBL-corrected spectrum using the EBL model of \citet{fin10}. For comparison, The EBL-corrected spectrum using the EBL model of \citet{fra08} is also shown as the red dotted curve. The observed data are taken from \citet{ice18b}.  }
\end{figure*}

\begin{figure*}
\centering
\includegraphics[width=12. cm,height=8. cm]{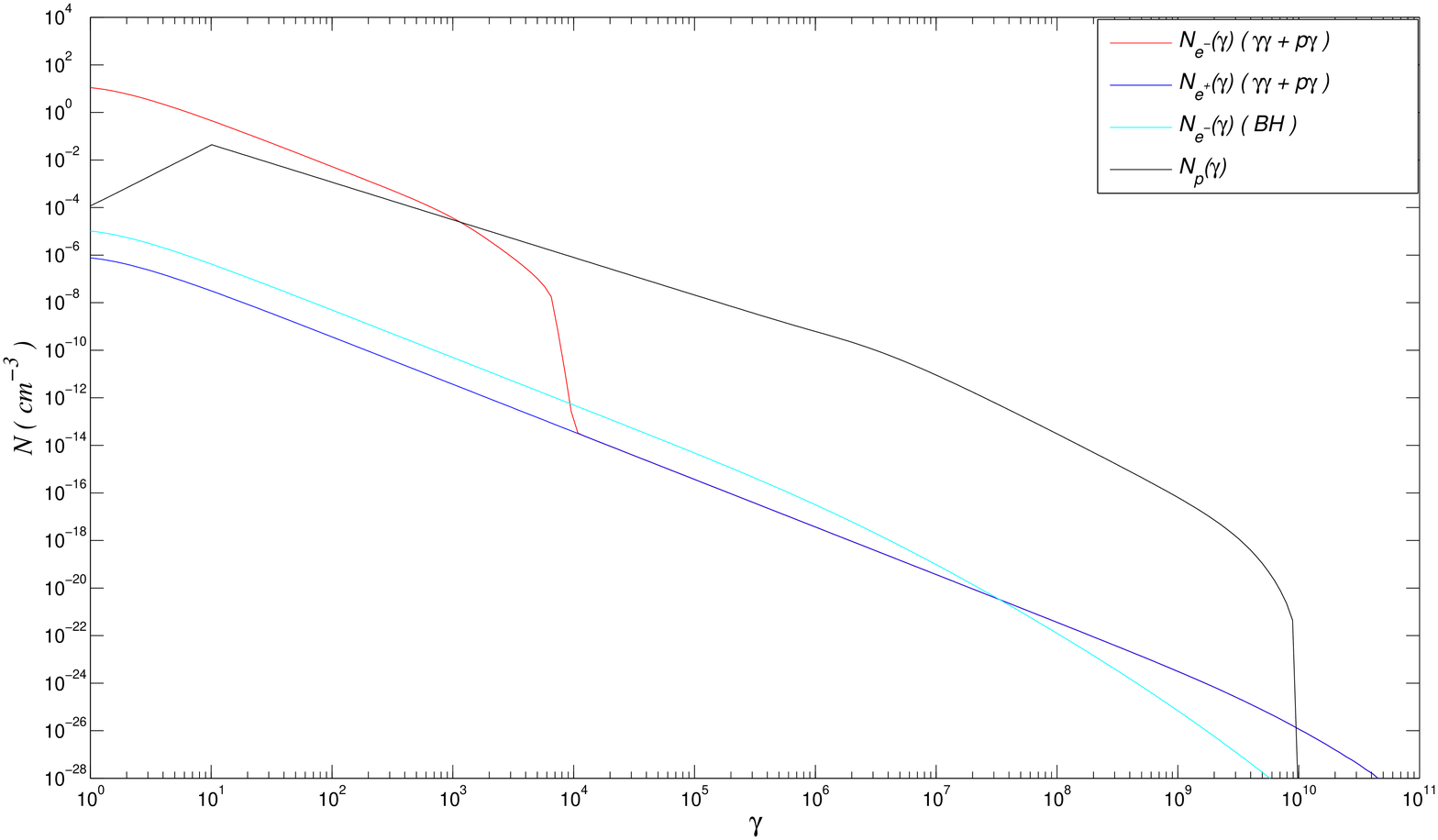}
\caption{Proton, electron and positron spectra from the derived SEDs in figure 1.}
\end{figure*}

\subsection{Kinetic equation in the acceleration zone  }

As the blob propagates through the jet, the particles from the upstream of jet are injected into the acceleration zone with an injected function
\begin{equation}
Q_{0.i}(\gamma)=Q_{0.i}\delta(\gamma-\gamma_{0,i}),
\end{equation}
where we assume a mono-energetic and time-independent injection. Each injected particle species are continuously accelerated up to very-high energies by stochastic acceleration processes with the synchrotron and SSC losses. The energy gain due to the acceleration is balanced by the radiative and escape losses. In the acceleration zone, a hard-sphere approximation is used to describe the plasma instabilities, hence the spatial diffusion coefficient $K_{\parallel,i}$ is independent of particle energy \citep{sch89,sta08}. This allows one to find the characteristic acceleration timescales due to stochastic acceleration  processes $t_{\rm acc,i}=v^2_{\rm A}/4K_{\parallel,i}$ \citep{wei10,wei15}, where $v_{\rm A}$ is Alfv\'{e}n speeds. We assume the acceleration timescale to be constant and scale linearly with the particle¡¯s mass. The timescale of the second species can be naturally obtained by relation $t_{\rm acc,i}\propto m_{\rm i}$. Moreover, the escape timescale is set to be constant and proportional to the acceleration timescale $t_{\rm esc,i} \propto t_{\rm acc,i}$. We expect that the stochastic acceleration  produces a power-law spectrum with spectral index $q\simeq1+t_{\rm acc}/(2 \, t_{\rm esc})$. This differs from the spectral index $q\simeq1+t_{\rm acc}/t_{\rm esc}$, which is expected from diffusive shock acceleration.
In fact, the stochastic acceleration produces a harder particle spectrum than the shock acceleration, which may be favorable to explain the observed SEDs from blazars (e.g., \cite{kat06,kak15,dil16,gao17}). In this paper,  we focus on the possibility of the stochastic acceleration in blazar jets.

We assume that the acceleration zone cannot directly produce the observed SEDs , hence we only solve the kinetic equations for the primary particles. The kinetic equations for primary electrons and protons are given by \citep{kat06}
\begin{eqnarray}
\label{acczone}
\partial_t n_i  = \partial_{\gamma} \big[( P_{s,i}(\gamma)+P_{IC,i}(\gamma) - t_{{acc,i}}^{-1}\gamma ) \cdot n_i\big]  \nonumber \\ \quad+ \partial_{\gamma} \big[ (2\,t_{{acc,i}})^{-1}\gamma^2 \partial_{\gamma} n_i\big] + Q_{0,i} - \frac{n_i}{t_{{esc,i}}}.
\end{eqnarray}
Where $P_{IC,i}(\gamma)$ is the inverse-Compton loss of each particle species, $P_{s,i}(\gamma)$ is the synchrotron loss of each particle species with
\begin{equation}
P_{s,i}(\gamma)=\frac{4}{3} \frac{ c \sigma_{\rm T}}{m_{\rm e} c^2}u_{B}\left (\frac{m_e}{m_{i}} \right )^3\gamma^2=\beta_{s,i}\gamma^2,
\end{equation}
where $\sigma_{\rm T}$ is the Thomson cross section, $u_{B}=\frac{B^2}{8\pi}$ is the energy density of the magnetic field.

\subsection{Kinetic equation in the radiation zone  }

As every escaping particle from the acceleration zone enters the radiation region, the particle spectrum $n_{i}(\gamma)$  from the acceleration zone severs as the injection function of the radiation zone. The particles are not accelerated in the radiation zone. Therefore, all relevant cool processes have to be taken into account, including synchrotron, inverse-Compton and photo-hadronic losses.
The kinetic equations for electrons, protons, photons, secondary positrons and neutrinos are solved self-consistently and time-dependently.

The low-energy photons from the primary electrons and protons are soft photons for the $p\gamma$ interactions. When the proton energy is above the threshold for the $p\gamma$ interactions, the high-energy protons will interact with the jet soft photons to produce secondary particles by the two channel:

The photon-pair prodution ( Bethe-Heitler process )
\begin{eqnarray}
p + \gamma \rightarrow e^{+} + e^{-} + p.
\end{eqnarray}
and the photon-meson production (see, e.g., \cite{rom08,kel08})
\begin{eqnarray}
 \label{chains}
 p+\gamma & \rightarrow & p + a\pi^{0}+b( \pi^{+}+\pi^{-} ) , \nonumber \\
 p+\gamma & \rightarrow & n + \pi^{+}+a\pi^{0}+b( \pi^{+}+\pi^{-} ).
\end{eqnarray}
Where $a$ and $b$ are the pion multiplicities. The produced pions are unstable particles, they decay into stable electrons, positrons, neutrinos and $\gamma$-ray photons by the channel:
\begin{eqnarray}
 \label{chains}
 \pi^{+} & \rightarrow & \mu^{+} + \nu_{\mu}, \,\,\, \mu^{+} \rightarrow e^{+} + \nu_e + \bar{\nu}_{\mu}, \nonumber \\
 \pi^{-} & \rightarrow & \mu^{-} + \bar{\nu}_{\mu}, \,\,\, \mu^{-}\rightarrow e^{-} + \bar{\nu}_e + \nu_{\mu}, \nonumber \\
 \pi^0 & \rightarrow & \gamma + \gamma.
\end{eqnarray}
This process will result in the third contribution to the VHE peak in the SEDs of blazars, besides the proton synchrotron emission and inverse-Compton emission from relativistic electrons. We use the Kelner \& Aharonian (2008) parametrization  of the SOPHIA Monte Carlo results to calculate the production rate of the stable electrons, positrons, neutrinos and $\gamma$-ray photons by the interactions of equation (6). We note that the decay timescale of unstable products from the $p\gamma$ interactions is shorter compared to the synchrotron loss timescale. Therefore, we do not account for the synchrotron losses and radiation of the intermediate particles ($\mu^{\pm}$, $\pi^{\pm}$).

\begin{table}
\caption{Model parameters of the proton synchrotron process} \label{para2}
\begin{center}
\begin{tabular}{lcccccccccccccccccccccccc}
\hline
& Parameters \\
\hline
&$B$ (G)                                       & & & &10                      \\
&$\delta_{\rm D}$                              & & & &48                      \\
&$\gamma_{0,e}$                                & & & &$6.0\times10^{3}$       \\
&$Q_{0,e}$ (cm$^{-3}$ s$^{-1}$)                & & & &$1.6\times10^{-1}$      \\
&$\gamma_{0,p}$                                & & & &10                      \\
&$Q_{0,p}$ (cm$^{-3}$ s$^{-1}$)                & & & &$4.0\times10^{-3}$      \\
&$R_{\rm acc}$ (cm)                            & & & &$3.0\times10^{13}$      \\
&$R_{\rm rad}$ (cm)                            & & & &$1.6\times10^{16}$      \\
&$t_{\rm acc}/t_{\rm esc}$                     & & & &$1.05$                  \\
\hline
\end{tabular}
\end{center}
\end{table}

The proton kinetic equation in the radiation zone is given by
\begin{eqnarray}
\label{radpro}
\partial_t N_{p}  &=&  \partial_{\gamma} \left[\big(P_{s,p}(\gamma) + P_{\rm{p}\gamma}(\gamma)+P_{\rm BH}(\gamma)\big) N_{p}\right] \nonumber
\\&&+ b \frac{n_{p}}{t_{esc,p}} - \frac{N_{p}}{t_{esc,rad,p}},
\end{eqnarray}
where  $b=(R_{\rm acc}/R_{\rm rad})^3<1$ is a geometric factor ensuring particle conservation. We use the formula given by \citet{kel08} to calculate the proton losses  due to the photon-meson interaction $P_{\rm{p}\gamma}(\gamma)$. The proton losses from the Bethe-Heitler process $P_{\rm BH}(\gamma)$ is calculated using the formula given by \citet{beg91}. As in the acceleration zone, the escape timescale is assumed to scale with the particle's mass, e.g. $t_{\rm esc,rad,i}\propto m_{\rm i}$, derived
from the particle's light crossing time of a sphere with radius $R_{\rm rad}$, multiplied by a constant empirical factor $\eta=10$.

The $\gamma$-rays from the $\pi^{0}$ decay and the secondary $e^{\pm}$ synchrotron emission are partially in the optically thick regime. The high-energy $\gamma$-ray photons can interacts with the low-energy synchrotron photons to produce the secondary pairs by
\begin{eqnarray}
\gamma+\gamma & \rightarrow &  e^{+} + e^{-}.
\end{eqnarray}
This process will initiate an electromagnetic cascade until the radiation enters the optically thin regime. The secondary pairs from the $\gamma$$\gamma$ interactions serve as an additional injection term in the kinetic equations of electron and positron.

The kinetic equations for the electrons and positrons in the radiation zone are thus given by
\begin{eqnarray}
 \partial_t N_{e^-} = \partial_{\gamma}\left[\big(P_{s,e}(\gamma)  + P_{{IC}}(\gamma)\big) \cdot N_{e^-} \right] - \frac{N_{e^-}}{t_{{rad,esc,e}}} \nonumber \\+ Q_{{\gamma\gamma}}(\gamma) + Q_{{p}\gamma^-}(\gamma)+Q_{\rm BH}(\gamma) + b\frac{n_{e^-}}{t_{{esc,e}}}, \label{radele}\\
 \partial_t N_{e^+} = \partial_{\gamma}\left[\big(P_{s,e}(\gamma)  + P_{{IC}}(\gamma)\big) \cdot N_{e^+} \right] - \frac{N_{e^+}}{t_{{rad,esc,e}}} \nonumber \\ + Q_{{\gamma\gamma}}(\gamma) + Q_{{p}\gamma^+}(\gamma)+Q_{\rm BH}(\gamma)\qquad\qquad.\label{radpos}
\end{eqnarray}
Note that no accelerated primary positrons are assumed in the model ($n_{e^{+}}=0$). The inverse-Compton loss rate $P_{{IC}}$ is calculated using the full Klein-Nishina cross section given
by \citet{blu70}. The secondary pair production rate $Q_{{p}\gamma}$ from the $p\gamma$ interaction is calculated using the $\Phi_{\pm}$-parameters of  the full SOPHIA Monte Carlo calculations  carried out by \citet{kel08}. The secondary pair production $Q_{{\gamma\gamma}}$ from the $\gamma\gamma$ interactions is calculated using the approximation of \citet{aha83}.
The photon-pair production rate $Q_{\rm BH}$ is calculated using the exact result of \citet{kel08}.

The kinetic equation for the photon field in the radiation zone is given by
\begin{eqnarray}
 \partial_t N_{\rm ph}  =  R_s(\nu) + R_c(\nu) + R_{\pi^0}(\nu) \qquad\qquad\quad\nonumber \\  - c \big( \alpha_{{SSA}}(\nu) + \alpha_{{\gamma\gamma}}(\nu) \big) N_{\rm ph}
  - \frac{N_{\rm ph}}{t_{{\rm esc,ph}}}\label{radpho},
\end{eqnarray}
where $t_{\rm esc,ph}=4R_{\rm rad}/3c$ is the photon escape timescale. The synchrotron production rate $R_{\rm s}$ and the absorption coefficient $\alpha_{\rm SSA}$ is calculated using the exact formula given by \citet{fin08}. The inverse-Compton production rate $R_c$ is calculated using the full Klein-Nishina cross section given by \citet{blu70}. The $\gamma$-ray production rate  from the $\pi^{0}$ decay, $R_{\pi^0}$, is calculated using $\Phi_{\gamma}$-parameters of \citet{kel08}. The photon annihilation coefficient $\alpha_{\gamma\gamma}$ is calculated using the exact result of \citet{cop90}.

\begin{figure*}
\centering
\includegraphics[width=13. cm,height=8. cm]{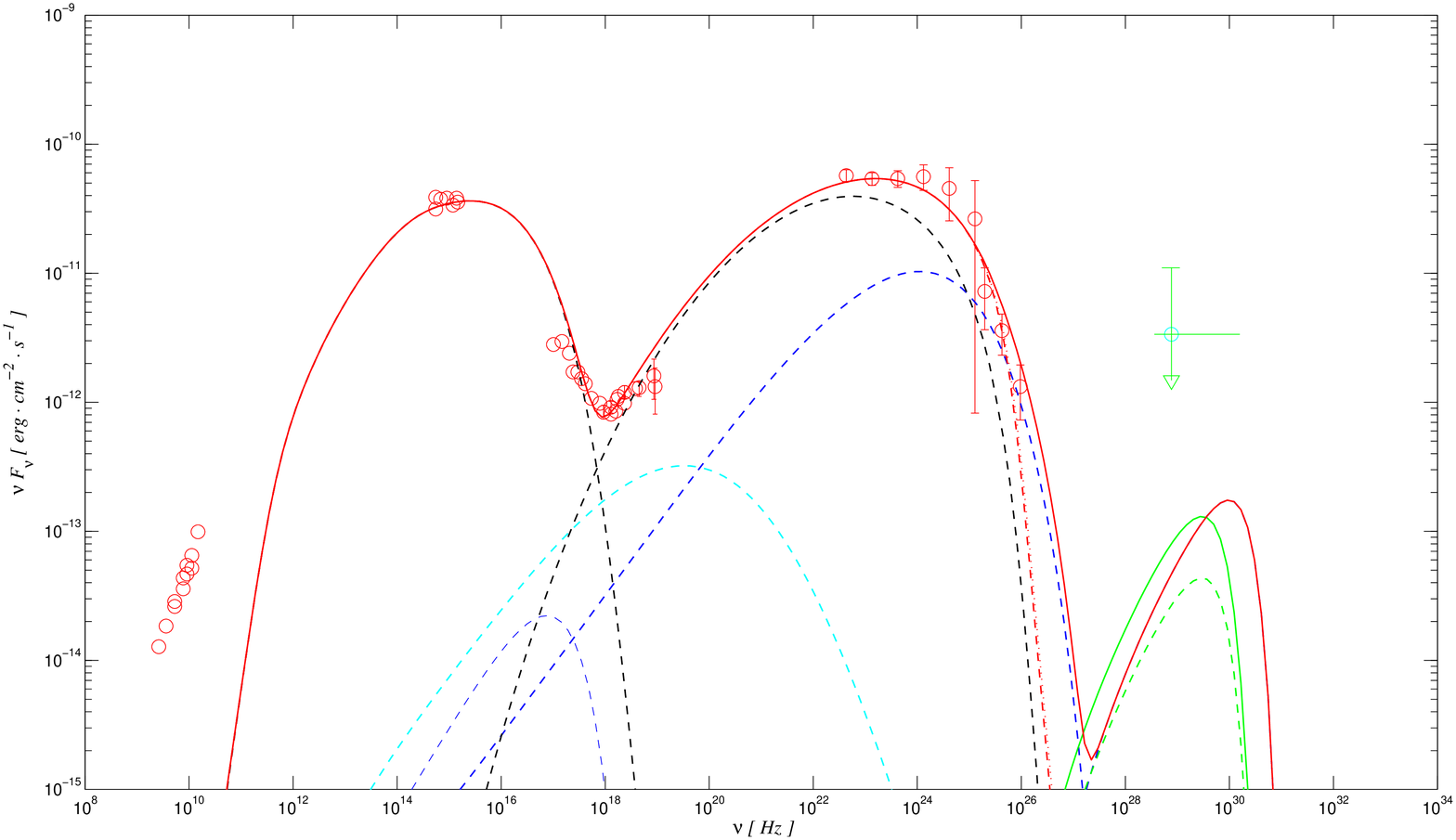}
\caption{Same as Figure 1, but for the hybrid leptonic-hadronic model.  }
\end{figure*}

\begin{figure*}
\centering
\includegraphics[width=12. cm,height=8. cm]{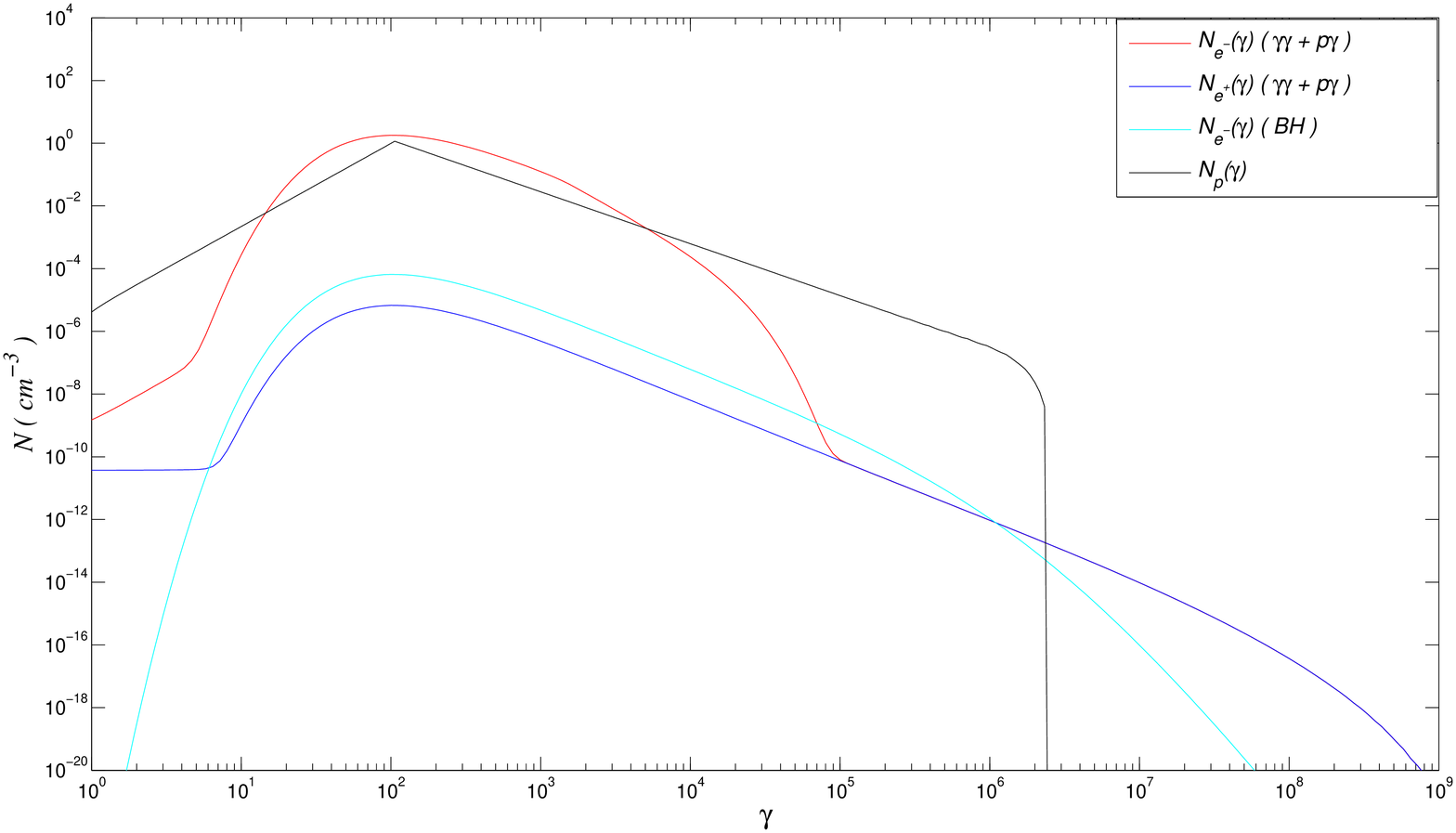}
\caption{Proton, electron and positron spectra from the derived SEDs in figure 3.  }
\end{figure*}

The neutrinos are not subject to any interaction except their production and escape. Therefore, the kinetic equation for the neutrinos are given by
\begin{eqnarray}
 \partial_t N_{\nu} & = & R_{\nu}(\nu)  - \frac{N_{\nu}}{t_{{\rm esc,ph}}}\label{radpho},
\end{eqnarray}
where the production rates for all favor neutrinos, $R_{\nu}(\nu)$, are calculated using the results of \citet{kel08}.

To model the observed SEDs of blazar, we need to transform the SEDs from the blob frame to that in the observer frame. The SEDs in the observer frame are given by
\begin{eqnarray}
 \nu F^{\rm obs}_{\nu}( \nu^{\rm obs}, t^{\rm obs} )=\frac{h \cdot \nu^2 \cdot N_{\rm ph}(\nu,t) \cdot \delta^4 \cdot V_{b}}{4\pi d^2_{\rm L} \cdot t_{\rm esc,ph}}
\end{eqnarray}
with $\nu^{\rm obs}=\delta  \nu$ and  $\triangle t^{\rm obs}=\triangle t / \delta $.

\subsection{Numeric method}
To obtain the model SEDs, we numerically solve a set of the coupled kinetic equations in the acceleration and radiation zone. In the acceleration zone, we use the method of \citet{cha70} to solve
the equations (3). In the radiation zone, we use the method of \citet{chi99} to solve the equations (7),\,(9) and (10). The  equations (11) and (12) are solved using the Crank-Nicolson method \citep{cra96}. We carefully tested our numerical code with some analytical solutions and found very good agreement.

\section{Result}
The blazar TXS 0506+056 is a bright BL Lac objects. The redshift of the source was recently measured to be $z=0.337$ \citep{pai18}. In September 2017, the IceCube reported a very-high-energy muon neutrino event (IceCube-170922A ), which was identified by the Extremely High Energy track event selection. The best-fit reconstructed direction is 0.1$^\circ$ from the sky position of the BL Lac object TXS 0506+056.  The energy of the neutrino event is estimated to be 290 TeV with the 90\% confidence level lower limits of 183 TeV  and  upper limits of 4.3 PeV, by assuming a power-law neutrino spectrum with the spectral index of -2.  The blazar TXS 0506+056 is a $\gamma$-ray source included in the third Fermi-LAT catalog of sources (Acero et al. 2015). Following the IceCuble alert, the Fermi-LAT reported that the direction of IceCube-170922A is coincident with the location of TXS 0506+056 and coincident with a state of the  $\gamma$-ray flare \citep{tan17}. The follow-up observations were performed by a multi-wavelength campaign with different telescopes, including a significant detection by MAGIC telescopes at $>$ 100 GeV, X-ray emissions by Swift/XRT and NuSTAR, optical emissions by the ASAS-SN survey as well as emission in radio band by VLA (see IceCube Collaboration 2018b ).
The high-energy neutrino originates in the hadronic interactions providing a natural link between high-energy $\gamma$-rays and neutrino. The combined multi-wavelength and neutrino observations provide a unique opportunity to study the hadronic processes in blazar jets.
\begin{table}
\caption{Model parameters of the leptonic-hadronic process} \label{para2}
\begin{center}
\begin{tabular}{lcccccccccccccccccccccccc}
\hline
& Parameters \\
\hline
&$B$ (G)                                       & & & &0.95                      \\
&$\delta_{\rm D}$                              & & & &28                      \\
&$\gamma_{0,e}$                                & & & &$1.5\times10^{3}$       \\
&$Q_{0,e}$ (cm$^{-3}$ s$^{-1}$)                & & & &$1.8\times10^{-1}$      \\
&$\gamma_{0,p}$                                & & & &$10^2$                  \\
&$Q_{0,p}$ (cm$^{-3}$ s$^{-1}$)                & & & &$1.4\times10^{-3}$      \\
&$R_{\rm acc}$ (cm)                            & & & &$1.0\times10^{14}$      \\
&$R_{\rm rad}$ (cm)                            & & & &$5.5\times10^{15}$      \\
&$t_{\rm acc}/t_{\rm esc}$                     & & & &$1.2$                  \\
\hline
\end{tabular}
\end{center}
\end{table}

\begin{figure*}
\centering
\includegraphics[width=13. cm,height=8. cm]{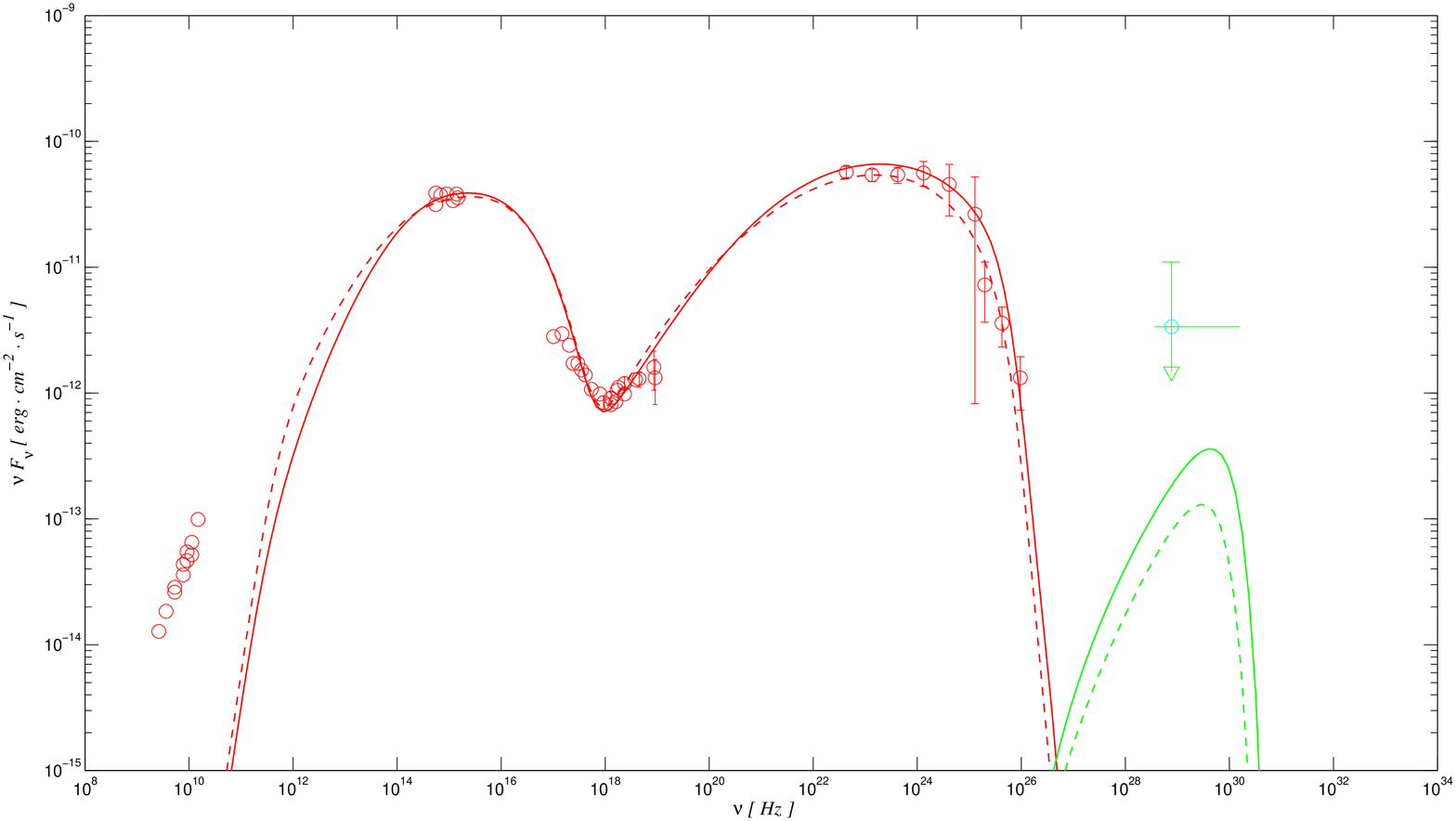}
\caption{Same as Figure 3, but for different Doppler factor. The solid curves represent the predicted electromagnetic and neutrino spectrum for the high Doppler factor ($\delta=40$). For comparison, the predicted electromagnetic and neutrino spectra for the low Doppler factor ($\delta=28$) are also shown as the dashed curves. }
\end{figure*}

We use the model described in section 2 to model the electromagnetic and neutrino spectra of blazar TXS 0506+056. The observed data is taken from \citet{ice18b}. We interpret the electromagnetic and neutrino spectra of blazar TXS 0506+056 by proton synchrotron and hybrid leptonic-hadronic models.
In figure 1, we show the predicted electromagnetic and neutrino spectra for the proton synchrotron model. The modeling parameters are listed in table 1.
The derived SEDs are corrected for the EBL absorption using the model of \citet{fra08} and \citet{fin10}. The different EBL models predict a similar $\gamma$-ray spectrum and the differences between  the two EBL models are negligible.
It can be seen that the optical and Swift X-ray spectra are produced by the synchrotron radiation of the primary electron, the NuSTAR X-ray spectrum comes from the low-energy tail of proton synchrotron radiation, the high-energy $\gamma$-ray spectrum is dominated by proton synchrotron radiation.
In fact, the contribution of the secondary emission from the pair cascades to the $\gamma$-ray spectrum is negligible due to strong EBL absorption above $10^{26}$ Hz. We note that the SSC emission from the primary electrons and the synchrotron emission of the secondary pairs from the Bethe-Heitler process have a negligible contribution to the observed SEDs due to high magnetic field of $\mathcal{O}$(10 G) (see table 1).
The peak energy of the predicted neutrino spectrum is $\sim 10^{32}$ Hz, which is far above the observed neutrino energy. In figure 2, we show the proton, electron and positron spectra from the derived SEDs in figure 1. In the acceleration zone, the stochastic acceleration process produces a power-law proton injected spectrum with index $q_{p}\simeq 1+t_{\rm acc,p}/(2\,t_{\rm esc,p})=1.525$.
In the radiation zone, the proton synchrotron loss dominates over the photo-meson loss due to high magnetic field. Therefore, the  synchrotron cooling results in a power-law proton spectrum with index $n_p\simeq q_p+1=2.525$ above the break energy $\gamma_{\rm b,p}\simeq1/(t_{\rm esc,rad,p}\,\beta_{\rm s,p})=5\times10^6$. The maximum proton energy can be determined by $\gamma_{\rm max,p}\simeq1/(t_{\rm esc,p}\,\beta_{\rm s,p})=2\times10^9$. The peak energy of proton synchrotron radiation is $\nu^{\rm obs}_{\rm s,p}\simeq4.2\times10^6 \delta \, B \, (m_{\rm e}/m_{\rm p}) \,\gamma^2_{\rm max,p}=5\times10^{24}$ Hz, which is comparable with the value from the model-derived SED (see figure 1). The electron density with $\gamma>10^{4}$ is the contribution of the secondary electrons from the pair cascade processes, because the primary electrons can not be accelerated to such high energies.
We note that the X-ray and $\gamma$-ray data can not be well reproduced  by the proton synchrotron model.

In figure 3, we show the predicted electromagnetic and neutrino spectra for the hybrid leptonic-hadronic model. In figure 4, we show the proton, electron and positron spectra from the derived SEDs in figure 3. The modeling parameters are listed in table 2. It can be seen that the optical and Swift X-ray spectra come from the synchrotron radiation of the primary electron.
The NuSTAR X-ray spectrum is produced by the combination of the SSC radiation and the synchrotron radiations of the secondary pairs from the Bethe-Heitler processes.
The high-energy $\gamma$-ray spectrum comes from the combination of the SSC radiation and the synchrotron radiation from the secondary pairs. We note that the synchrotron radiation from the secondary pairs  mainly contributes to the high-energy tails of the $\gamma$-ray spectrum. In fact, the proton synchrotron radiation makes a negligible contribution to the observed SEDs due to low magnetic field of $\mathcal{O}$(1 G) (see table 2).
The neutrino peak energy can be estimated by the maximum proton energy from the relation of $E_{\nu}=0.05 \, E_{\rm max,p}$. Therefore, the corresponding neutrino peak frequency is $\nu^{\rm obs}_{\nu}\simeq 3\times10^{29}(\gamma_{\rm max,p}/10^6)(\delta/28)$ Hz, which is closer to the observed neutrino energy compared to the proton synchrotron emission.
Recently,  the study of MAGIC collaboration showed that the blazar TXS 0506+056 has a Doppler factor about 40 \citep{ans18}.  We also show the predicted electromagnetic and neutrino spectra for the high Doppler factor ($\delta=40$) as the  solid curves in figure 5. The modeling parameters are listed in table 3. For comparison, the predicted electromagnetic and neutrino spectra for the low Doppler factor ($\delta=28$) are also shown as the dashed curves.  It is found that the model of the high Doppler factor predicts a slightly higher neutrino flux than one of the low Doppler factor. Our results inmpiled that the hybrid leptonic-hadronic model can better match the X-ray and $\gamma$-ray spectra than the proton synchrotron model (see figure 1 and 5).

\begin{table}
\caption{Model parameters of the leptonic-hadronic process for the high Doppler factor } \label{para2}
\begin{center}
\begin{tabular}{lcccccccccccccccccccccccc}
\hline
& Parameters \\
\hline
&$B$ (G)                                       & & & &1                      \\
&$\delta_{\rm D}$                              & & & &40                      \\
&$\gamma_{0,e}$                                & & & &$1.5\times10^{3}$       \\
&$Q_{0,e}$ (cm$^{-3}$ s$^{-1}$)                & & & &$5.6\times10^{-2}$      \\
&$\gamma_{0,p}$                                & & & &$10^2$                  \\
&$Q_{0,p}$ (cm$^{-3}$ s$^{-1}$)                & & & &$2.3\times10^{-3}$      \\
&$R_{\rm acc}$ (cm)                            & & & &$1.0\times10^{14}$      \\
&$R_{\rm rad}$ (cm)                            & & & &$2.5\times10^{15}$      \\
&$t_{\rm acc}/t_{\rm esc}$                     & & & &$1.3$                  \\
\hline
\end{tabular}
\end{center}
\end{table}
\section{Discussion and Conclusions}
The $pp$ interactions have been invoked to explain the observed neutrino spectrum of Blazar TXS 0506+056 \citep{liu18,wan18,he18,sah18}.
The $pp$ interaction usually requires the high plasma density which is not expected in the environment of BL Lac objects due to small accretion rate\citep{aha00}. The observed neutrino spectrum can be explained by the $p\gamma$ interactions with an external radiation
field \citep{kei18}. It is thought that BL Lac objects are lack of a strong external radiation field due to the clean environment around the jet.
A steady leptonic-hadronic model was used to study the electromagnetic and neutrino emissions of blazar TXS 0506+056\citep{cer18,zha18}. However, these model do not include the particle acceleration and the relevant radiative processes self-consistently.

In this paper, we study the correlated electromagnetic and neutrino emission of blazar TXS 0506+056 by a self-consistent leptonic-hadronic model, taking into account both electron and proton acceleration and all relevant radiative processes self-consistently. We reproduce the electromagnetic and neutrino spectra of blazar TXS 0506+056 by the proton synchrotron and hybrid leptonic-hadronic models based on the $p\gamma$ interaction. We find that the hybrid leptonic-hadronic model can better reproduce the observed multi-wavelength SEDs of blazar TXS 0506+056. Moreover, the predicted neutrino spectrum of the hybrid leptonic-hadronic model is closer to the observed one compared to the proton synchrotron model.
Therefore, we suggest that the hybrid leptonic-hadronic model is more preferred if the neutrino IceCube-170922A originates from blazar TXS 0506+056.
It is not possible to put a
strong constrain on neutrino production models with single observed neutrino event. Further multi-messenger observations will be needed to understand the possible physical connection between the
neutrino and $\gamma$-ray emissions.

\begin{ack}
 We thank the anonymous referee for valuable comments and suggestions. We acknowledge the financial support from the National Natural Science Foundation of China 11573060 and 11663008,
the National Science Foundation of China 11673060,  and the Natural Science Foundation of Yunnan Province under grant 2016FB003.

\end{ack}

%
%


\end{document}